# Network Coding for Distributed Storage Systems

Alexandros G. Dimakis, P. Brighten Godfrey, Martin J. Wainwright and Kannan Ramchandran
Department of Electrical Engineering and Computer Science,
University of California, Berkeley, CA 94704.
Email: {adim, pbg, wainwrig, kannanr}@eecs.berkeley.edu

***Abstract*— Peer-to-peer distributed storage systems provide reliable access to data through redundancy spread over nodes across the Internet. A key goal is to minimize the amount of bandwidth used to maintain that redundancy. Storing a file using an erasure code, in fragments spread across nodes, promises to require less redundancy and hence less maintenance bandwidth than simple replication to provide the same level of reliability. However, since fragments must be periodically replaced as nodes fail, a key question is how to generate a new fragment in a distributed way while transferring as little data as possible across the network.**

**In this paper, we introduce a general technique to analyze storage architectures that combine any form of coding and replication, as well as presenting two new schemes for maintaining redundancy using erasure codes. First, we show how to optimally generate MDS fragments directly from existing fragments in the system. Second, we introduce a new scheme called Regenerating Codes which use slightly larger fragments than MDS but have lower overall bandwidth use. We also show through simulation that in realistic environments, Regenerating Codes can reduce maintenance bandwidth use by $25\%$ or more compared with the best previous design—a hybrid of replication and erasure codes—while simplifying system architecture.**

## I. Introduction

The purpose of distributed file storage systems such as OceanStore [19], Total Recall [2], and DHash++ [6] is to store data reliably over long periods of time using a distributed collection of disks (say, at various nodes across the Internet). Ensuring reliability requires the introduction of redundancy, the simplest form of which is straightforward replication.

Several designs [18], [2], [6] use erasure codes instead of replication. A *Maximum-Distance Separable* (MDS) erasure code stores a file of size $\mathcal{M}$ bytes in the form of $n$ fragments each of size $\mathcal{M}/k$ bytes, any $k$ of which can be used to reconstruct the original file.

However, a complication arises: in distributed storage systems, redundancy must be continually refreshed as nodes choose to leave the system and disks fail, which involves large data transfers across the network. How do we efficiently create new encoded fragments in response to failures? A new replica may simply be copied from any other node storing one, but traditional erasure codes require access to the original data to produce a new encoded fragment. How do we generate an erasure encoded fragment when we only have access to erasure encoded fragments?

In the *naive strategy*, the node which will store the new fragment—which we will call the *newcomer*—downloads $k$ fragments and reconstructs the file, from which a new fragment is produced. Thus, $\mathcal{M}$ bytes are transferred to generate a fragment of size only $\mathcal{M}/k$.

To reduce bandwidth use, one can adopt what we call the *Hybrid strategy* [20]: one full replica is maintained in addition to multiple erasure-coded fragments. The node storing the replica can produce new fragments and send them to newcomers, thus transferring just $\mathcal{M}/k$ bytes for a new fragment. However, maintaining an extra replica on one node dilutes the bandwidth-efficiency of erasure codes and complicates system design. For example, if the replica is lost, new fragments cannot be created until it is restored. In fact, one study comparing the Hybrid strategy with replication in distributed storage systems [20] argued that in practical environments, Hybrid's reduced bandwidth is limited, and may be outweighed by its drawbacks, in part due to the added complication of maintaining two types of redundancy.

It is thus natural to pose the following question: is it possible to maintain an erasure code using less bandwidth than the naive strategy, without resorting to an asymmetric strategy like Hybrid? More deeply, what is the minimal amount of data that must be downloaded in order to maintain an erasure code?

In this paper we show how network coding can help for such distributed storage scenarios. We introduce a general graph-theoretic framework through which we obtain lower bounds on the bandwidth required to maintain any distributed storage architecture and show how random linear network coding can achieve these lower bounds.

More specifically, we determine the minimum amount of data that a newcomer has to download to generate an MDS or nearly-MDS fragment, a scheme which we call *Optimally Maintained MDS* (OMMDS). In particular, we prove that if the newcomer can only connect to $k$ nodes to download data for its new fragment, then the $\mathcal{M}$-byte download of the naive strategy is the information-theoretic minimum. Surprisingly, if the newcomer is allowed to connect to more than $k$ nodes, then the total download requirement can be reduced significantly. For example, if $k = 7$ (the value used in DHash++ [6]), $n = 14$, and a newcomer connects to $n-1$ nodes, a new fragment can be generated by transferring $0.27\mathcal{M}$ bytes, or 73% less than the naive strategy. However, the associated overhead is still substantial, and it turns out that Hybrid offers a better reliability-bandwidth tradeoff than OMMDS. To improve on Hybrid, we must therefore look beyond MDS codes.

With this perspective in mind, we introduce our second scheme, *Regenerating Codes* (RC), which minimize amount of data that a newcomer must download subject to the restriction that we preserve the "symmetry" of MDS codes. At a high level, the RC scheme improves on OMMDS by having a newcomer store all the data that it downloads, rather than throwing some away. As a consequence, RC has slightly *larger* fragments than MDS, but very low maintenance bandwidth overhead, even when newcomers connect to just $k$ nodes. For example, if $k = 7$, a newcomer needs to download only $0.16\mathcal{M}$ bytes—39% less than OMMDS and 84% less than the naive strategy. Moreover, our simulation results based on measurements of node availability in real distributed systems show that RC can reduce bandwidth use by up to 25% compared with Hybrid when $k = 7$. RC improves even further as $k$ grows.

We emphasize that there are still tradeoffs between RC and other strategies. For example, users wishing to reconstruct the file pay a small overhead due to RC's larger fragments. Nevertheless, RC offers a promising alternative due to its simplicity and low maintenance bandwidth.

In summary, the contributions of this paper are as follows.

- We introduce a framework to analyze the bandwidth requirements of redundancy schemes for distributed storage systems.
- We characterize the minimum bandwidth necessary to produce an MDS fragment directly from fragments on other nodes.
- We introduce a non-MDS scheme, Regenerating Codes, and show through simulation that it requires substantially lower maintenance bandwidth than the best previous erasure code-based scheme (Hybrid) while preserving the symmetry of MDS codes.

This paper is organized as follows. We discuss relevant background and related work from coding theory and distributed storage systems in Section II. In Section III we introduce our analysis technique and use it to determine how to optimally maintain MDS codes in Section III-B. We introduce Regenerating Codes in Section IV. Finally, Section V compares Hybrid, OMMDS, and RC using measured traces of node availability and discusses qualitative tradeoffs between the strategies.

## II. Background and Related Work

### A. Erasure codes

Erasure Coding is a generalization of replication that divides the initial data object into $k$ fragments (or blocks) which are then used to generate $n$ encoded fragments. MDS (Maximum-Distance Separable) erasure codes have the property that *any* $k$ (or slightly more) out of the $n$ encoded fragments suffice to recover the original $k$ data fragments. Good (i.e MDS or nearly MDS) erasure codes yield much higher probabilities of recovery compared to replication schemes but also introduce higher computational complexity. One way to theoretically quantify that benefit is the coupon collector problem: It is necessary to obtain $k \ln k$ randomly selected fragments to collect all $k$ original data, and in that sense erasure coding saves an $\ln k$ factor. Reducing encoding and decoding complexity for erasure codes has been studied extensively, and currently essentially optimal erasure codes exist with linear encoding and decoding complexity [17], [23]. Fountain codes [16], [23] (also called rateless codes) allow the creation of each encoded fragment independently and are therefore useful for many scenarios, such as distributed storage systems, which need to create new fragments continuously as nodes join and leave the system.

### B. Network Coding

Ahlswede et al [1] introduced the fundamental idea of Network Coding—combining packets instead of just routing—and showed that it achieves the minimum of the min-cuts for multicasting Later it was shown that linear operations over finite fields are sufficient [15] to achieve the network coding capacity. See [9] for an up-to-date survey of the area.

For distributed storage, the idea of using network coding was introduced in [8] in a sensor network scenario. Many aspects of coding for storage were further explored [14], [30], [26] for sensor network applications.

Network coding was proposed for peer-to-peer content distribution systems [10] where random linear operations over packets are performed to improve downloading. Random network coding was also recently proposed for P2P network diagnosis [29]. Our paper is based on similar ideas but the storage systems have different performance metrics that need to be analyzed.

### C. Distributed storage systems

A number of recent studies [4], [18], [7], [21], [2], [27] have designed and evaluated large-scale, peer-to-peer distributed storage systems. Redundancy management strategies for such systems have been evaluated in [28], [3], [2], [20], [27], [5], [25], [11].

Among those, [28], [2], [20] compared replication with erasure codes in the bandwidth-reliability tradeoff space. The analysis of Weatherspoon and Kubiatowicz [28] showed that erasure codes could reduce bandwidth use by an order of magnitude compared with replication. Bhagwan et al [2] came to a similar conclusion in a simulation of the Total Recall storage system.

Rodrigues and Liskov [20] arrived at a different result: in high-churn (i.e., high rate of node turnover) environments, erasure codes provide a large benefit but the bandwidth cost is too high to be practical for a P2P distributed storage system. In low-churn environments, the reduction in bandwidth is negligible. In moderate-churn environments, there is some benefit, but this may be outweighted by the added architectural complexity that erasure codes introduce (discussed further in Section V-E). These conclusions were based on an analytical model augmented with parameters estimated from traces of real systems. Compared with [28], [20] used a much smaller value of $k$ (7 instead of 32) and the Hybrid strategy to address the code regeneration problem.

In Section V, we repeat the evaluation of [20] to measure the performance of the two redundancy maintenance schemes that we introduce.

## III. Fundamental limits on bandwidth

### A. Information flow graph

Our analysis is based on a particular graphical representation of a distributed storage system, which we refer to as an *information flow graph* $\mathcal{G}$. This graph describes how the information of the data object travels through time and storage nodes and reaches reconstruction points at the data collectors. More precisely, it is a directed acyclic graph consisting of three kinds of nodes: a single data source $\mathsf{S}$, storage nodes $\mathsf{x_{in}}^i, \mathsf{x_{out}}^i$ and data collectors $\mathsf{DC}_i$. The single node $\mathsf{S}$ corresponds to the source of the original data. Storage node $i$ in the system is represented by a storage input node $\mathsf{x_{in}}^i$, and a storage output node $\mathsf{x_{out}}$; these two nodes are connected by a directed edge $\mathsf{x_{in}}^i \rightarrow \mathsf{x_{out}}^i$ with capacity equal to the amount of data stored at node $i$. See Figure III-A for an illustration.

Given the dynamic nature of the storage systems that we consider, the information flow graph also evolves in time. At any given time, each vertex in the graph is either *active* or *inactive*, depending on whether it is available in the network. At the initial time, only the source node $\mathsf{S}$ is active; it then contacts an initial set of storage nodes, and connects to their inputs ($\mathsf{x_{in}}$) with directed edges of infinite capacity. From this point onwards, the original source node $\mathsf{S}$ becomes and remains inactive. At the next time step, the initially chosen storage nodes become now active; they represent a distributed erasure code, corresponding to the desired steady state of the system. If a new node $j$ joins the system, it can only be connected with active nodes. If the newcomer $j$ chooses to connect with active storage node $i$, then we add a directed edge from $\mathsf{x_{out}}^i$ to $\mathsf{x_{in}}^j$, with capacity equal to the amount of data that the newcomer downloads node $i$. Note that in general it is possible for nodes to download more data than they store, as in the example of the $(14, 7)$-erasure code. If a node leaves the system, it becomes inactive. Finally, a data collector $\mathsf{DC}$ is a node that corresponds to a request to reconstruct the data. Data collectors connect to subsets of active nodes through edges with infinite capacity.

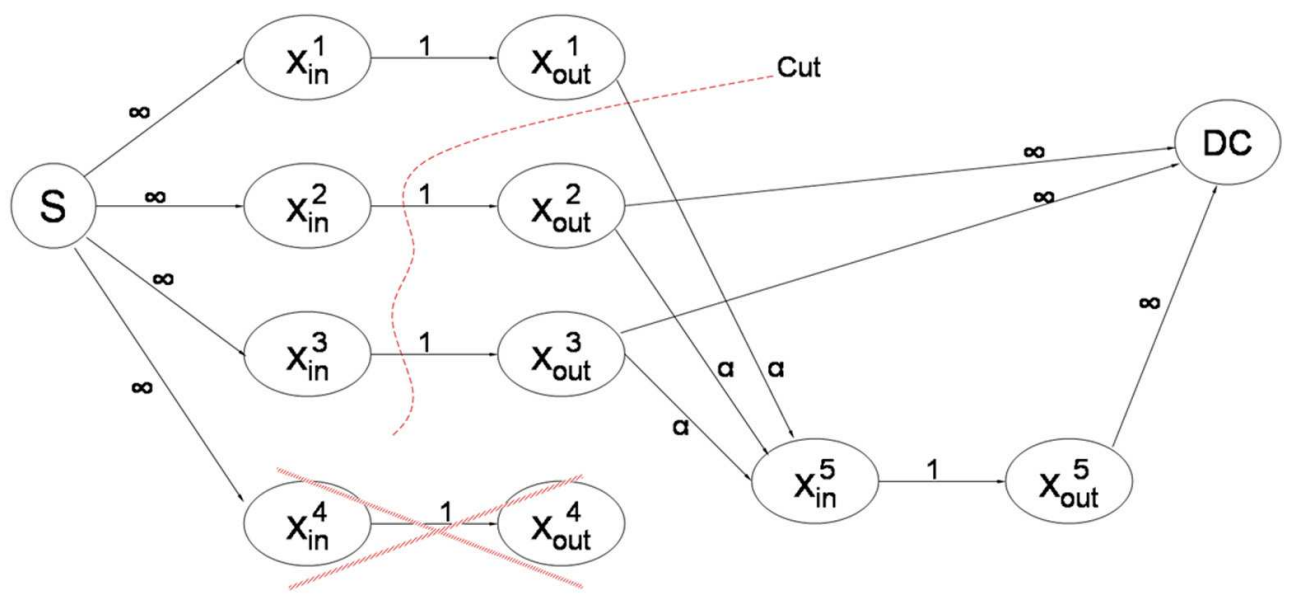

**Fig. 1.** Illustration of an information flow graph $\mathcal{G}$. Suppose that a particular distributed storage scheme uses an $(4, 3)$ erasure code in which any 3 fragments suffice to recover the original data. If node $x^4$ becomes unavailable and a new node joins the system, then we need to construct new encoded fragment in $x^5$. To do so, node $x^5_{in}$ is connected to the $k = 3$ active storage nodes. Assuming that it downloads $\alpha$ bits from each active storage node, of interest is the minimum $\alpha$ required. The min-cut separating the source and the data collector must be larger than 3 for reconstruction to be possible. For this graph, the min-cut value is given by $2 + \alpha$, implying that $\alpha \geq 1$, so that the newcomer has to download the complete data object if he connects to only $k = 3$ storage nodes.

An important notion associated with the information flow graph is that of minimum cuts:

*Definition 1:* A cut in the graph $\mathcal{G}$ between the source $\mathsf{S}$ and a fixed data collector node $\mathsf{DC}$ is a subset $C$ of edges such that, there is no path starting from $\mathsf{S}$ to $\mathsf{DC}$ that does not have one or more edges in $C$. The minimum cut is the cut between $\mathsf{S}$ and $\mathsf{DC}$ in which the total sum of the edge capactities is smallest.

### B. Bounds

To obtain bounds on the how much each storage node has to download, we start with the following simple lemma.

*Lemma 1:* A data collector $\mathsf{DC}$ can never reconstruct the initial data object if the minimum cut in $\mathcal{G}$ between $\mathsf{S}$ and $\mathsf{DC}$ is smaller than the initial object size.

*Proof:* The information of the initial data object is flowing from the source to the particular data collector. Every link in the information flow graph can only be used at most once (since it corresponds to communication of nodes over time), and since the point-to-point capacity is less than the file size, communication of the initial data object is impossible. ∎

The next claim, which builds on known results from network coding, shows that there exist linear network codes which can match this bound for *all data collectors simultaneously*, and also that simple linear mixing of packets using random independent coefficients over a finite field (randomized network coding [13]) will be sufficient with high probability.

*Proposition 1:* Assume that for some distributed storage scheme, we construct the $\mathcal{G}$ graph and place all the possible $\binom{n}{k}$ data collectors where $n$ is the number of active nodes. If the minimum of the min-cuts separating the source with each data collector is larger or equal to the data object size $\mathcal{M}$, then there exists a linear network code such that all data collectors can recover the data object. Further, randomized network coding guarantees that all collectors can recover the data object with probability that can be driven arbitrarily high by increasing the field size.

*Proof:* The reconstruction problem described is equivalent to the multicasting problem, with a single source sending its data to all of the data collectors. It is known [1] that network coding can achieve the associated min-cut/max-flow bound and from [15] we know that a linear network code will exist (see also Section II).

Ho et al. [13] show that the use of random linear network codes at all storage nodes suffices to achieve the bound with probability that can be pushed arbitrarily high by increasing the field size. (See in particular Theorem 3 in the paper [13], which ensures that the probability is at least $(1-\frac{d}{q})^N$, where $d$ is the number of data collectors and $N$ is total number of storage nodes in $\mathcal{G}$ and $q$ is the field size.) As in all the the work that uses network coding, the field size can be made very large easily since it is exponential in the number of bits used to represent field elements. ■

The above results allow us to provide a complete characterization of the bandwidth cost associated with maintaining an MDS erasure code:

*Proposition 2:* Assume the data object is divided in $k$ fragments, an $(n,k)$-MDS code is generated and one encoded fragment is stored at each node. Suppose a newcomer creates a new encoded fragment by downloading $\alpha$ percent of a fragment from each of $n-1$ active storage nodes. Then $\alpha \geq \frac{1}{n-k}$ is necessary and sufficient for successful reconstruction.

*Proof:* Consider the information flow graph $\mathcal{G}$ for this storage system. Suppose that any newcomer connects to $n-1$ storage nodes and downloads a portion $\alpha$ of the fragment from each storage node, where $\alpha$ is to be determined. A data collector is connected to the newcomer and $k-1$ other storage nodes. The minimum cut in this newly formed $\mathcal{G}$ is given by $k-1+(n-1-(k-1))\alpha$; so using proposition 1, successful reconstruction is possible if and only if this cut is larger or equal to $k$. So $\alpha \geq \frac{1}{n-k}$ is the minimum possible bandwidth to required maintain an MDS code. ■

Note that the information flow graph can be used to find the bandwidth requirements in the more general case where the newcomer connects to $h \leq n-1$ nodes and it is not hard to verify that when $h=k$ the whole file needs to be downloaded to create a new encoded fragment. In the special case of the $(n,k)=(14,7)$ erasure code considered in our motivating example, Proposition 2 verifies the earlier claim that the newcomer needs to download only $\frac{1}{7} \approx 0.14$Mb from each of the $n-1=13$ active storage nodes. We refer to MDS codes maintained in this procedure specified by Proposition 2 as *Optimally Maintained MDS*, or *OMMDS* for short.

## IV. Regenerating codes

The OMMDS scheme of the previous section is a significant improvement over the naive scheme of downloading the entire file to generate a new fragment. However, the associated overhead is still substantial, and our experimental evaluation in Section V reveals that the Hybrid scheme still offers a better reliability-bandwidth tradeoff than the OMMDS. Moreover, as established in Proposition 2, an MDS code cannot be maintained with less bandwidth than OMMDS. Therefore, we can only hope to use less bandwidth with a coding scheme other than an MDS code.

With this perspective in mind, this section introduces the notion of a *Regenerating Code* (RC). Subject to the restrictions that we preserve the "symmetry" of MDS codes (detailed in Section IV), we derive matching lower and upper bounds on the minimal amount of data that a newcomer must download. In contrast with OMMDS, the RC approach has very low bandwidth overhead, even when newcomers connect to just $k$ nodes. At a high level, the RC scheme improves on OMMDS by having a newcomer store *all* the data that it downloads, rather than throwing some away. As a consequence, RC fragments are slightly larger than MDS fragments, by a factor $\beta_{\text{RC}} = k^2/(k^2-k+1)$ (see Figure 2 for an illustration), and any data collector that reconstructs the file downloads $\beta_{\text{RC}}$ times the size of the file. However, note that $\beta_{\text{RC}} \to 1$ as $k \to \infty$. Notice that for MDS codes, if we fix the rate of the code $R = k/n$, the bandwidth overhead is $\beta_{\text{MDS}} = \frac{n-1}{n-k} \to \frac{1}{1-R}$ (which is constant) as $k,n \to \infty$. Therefore, MDS codes have a constant multiplicative overhead in bandwidth, but are optimal in storage for any $n,k$. The surprising fact is that regenerating codes, by sacrificing an (asymptotically) negligible factor $\beta_{\text{RC}}$ in storage, also achive asymptotically negliglible overhead in maintenance bandwidth.

Regenerating codes minimize the required bandwidth under a "symmetry" requirement over storage nodes. Specifically, we require that any $k$ fragments can reconstruct the original file; all fragments have equal size $\alpha\mathcal{M}$; and a newcomer produces a new fragment by connecting to any $k$ nodes and downloading $\alpha\mathcal{M}/k$ bits from each (In this paper we fix the number of nodes where the newcomer connects to to be $k$ (the minimum possible) to simplify the scheme). The free parameter $\alpha$ will be chosen to minimize bandwidth.

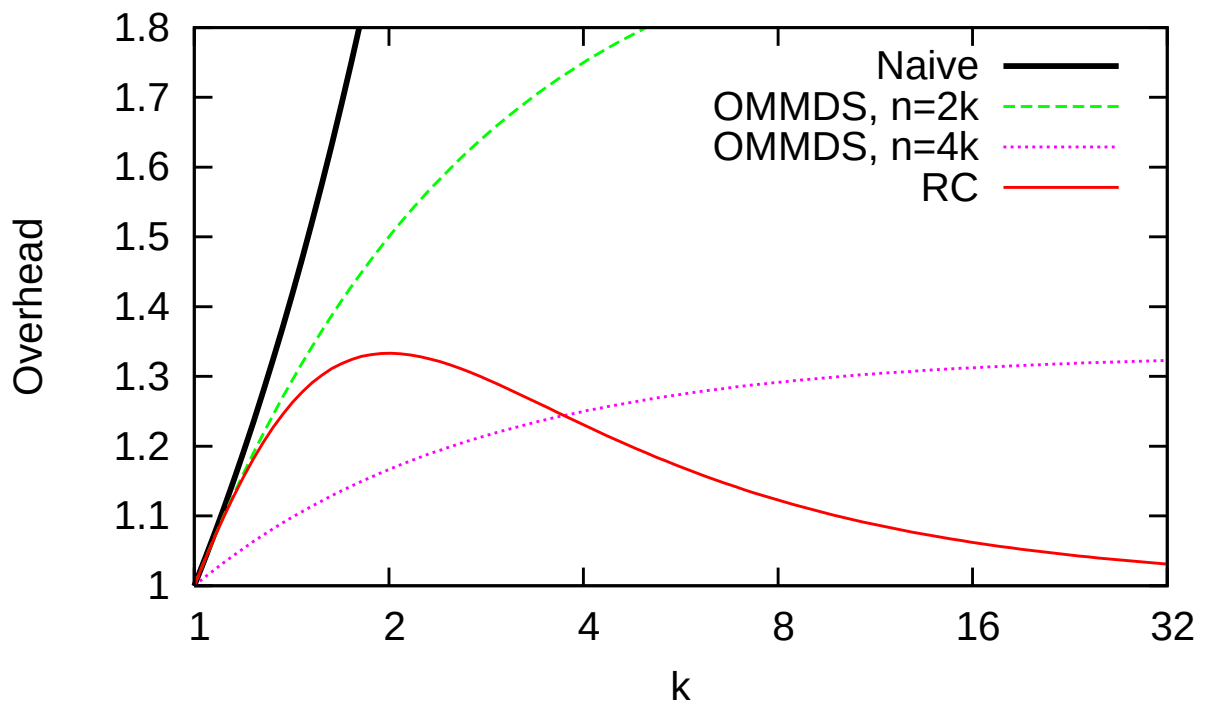

**Fig. 2.** The *overhead* $\beta$ is the number of bytes downloaded to produce a fragment, divided by the size of an MDS fragment. For the naive strategy, $\beta_{\text{naive}} = k$; for OMMDS in which newcomers connect to $n-1$ nodes, $\beta_{\text{MDS}} = \frac{n-1}{n-k}$; for RC in which newcomers connect to just $k$ nodes, $\beta_{\text{RC}} = k^2/(k^2-k+1)$. Moreover, RC fragments are $\beta_{\text{RC}}$ times larger than MDS fragments, so that the data collector must download $\beta_{\text{RC}}$ times the size of the original file.

Assume that newcomers arrive sequentially, and that each one connects to an arbitrary $k$-subset of previous nodes (including previous newcomers). The following result characterizes the bandwidth requirements of the RC scheme:

*Theorem 1:* Assume all storage nodes store $\alpha\mathcal{M}$ bits and newcomers connect to $k$ existing nodes and download $\frac{1}{k}\alpha\mathcal{M}$ bits from each. Then, define

$$\alpha_c = \frac{1}{k} \times \frac{1}{1 - \frac{1}{k} + \frac{1}{k^2}}. \tag{1}$$

If $\alpha < \alpha_c$ then reconstruction at some data collector who connects to $k$ storage nodes is information theoretically impossible.

If $\alpha \geq \alpha_c$ there exists a linear network code such that any data collector can reconstruct. Moreover, randomized network coding at the storage nodes will suffice with high probability.

*Proof:* We will show that if $\alpha < \alpha_c$ the minimum cut from some $k$ subset of storage nodes to the source S will be less than $\mathcal{M}$ and therefore reconstruction will be impossible. In addition when $\alpha \geq \alpha_c$ the minimum cut will be greater or equal to $\mathcal{M}$. Then by Proposition 1 a linear network code exists so that all data collectors can recover. Further randomized network coding will work with probability that can be driven arbitrarily high by increasing the field size.

Therefore it suffices to find the minimum $\alpha_c$ such that any $k$ subset of storage nodes has a minimum cut from the source equal to $\mathcal{M}$. We proceed via induction on $n$, the number of storage nodes. We refer to any subgraph of $\mathcal{G}$ with $k$ inputs and $j \geq k$ outputs as a *box*; a box is called *good* if every $k$ out of the $j$ outputs can support an end-to-end flow of $\mathcal{M}$. The base case of the induction is trivial if we assume that there are $k$ storage nodes initially.

For the inductive step, assume we have a good box denoted $B_{j-1}$ and a newcomer $X_i$ connects to any $k$ outputs of $B_{j-1}$ with edges that have capacity $\alpha\frac{\mathcal{M}}{k}$ (see figure IV). One needs to show that the new graph with the outputs of $B_{j-1}$ plus the output of the storage node $X_i$ will be a good box $B_j$. Let $N(X_i)$ denote the storage nodes where $X_i$ connected to. Consider a data collector that connects to $y_1$ nodes in $N(X_i)^c$ and $y_2$ nodes in $N(X_i)$, and also to the newcomer (all data collectors that do not connect to the newcomer receive enough flow by the induction hypothesis). We therefore have $y_1 + y_2 = k-1$ and also the minimum cut for this data collector is

$$y_1\alpha\mathcal{M} + y_2\alpha\mathcal{M} + (k-y_2)\frac{\alpha\mathcal{M}}{k}. \tag{2}$$

To ensure recovery this has to work for every data collector, i.e.

$$y_1\alpha\mathcal{M} + y_2\alpha\mathcal{M} + (k-y_2)\frac{\alpha\mathcal{M}}{k} \geq \mathcal{M}, \tag{3}$$

$$\forall y_1, y_2, y_1 + y_2 = k-1. \tag{4}$$

It is easy to see that $y_1 = 0$ is the worst case, and from there one obtains that

$$\alpha \geq \frac{1}{k(1-\frac{1}{k}+\frac{1}{k^2})} =: \alpha_c \tag{5}$$

is necessary and sufficient for reconstruction.

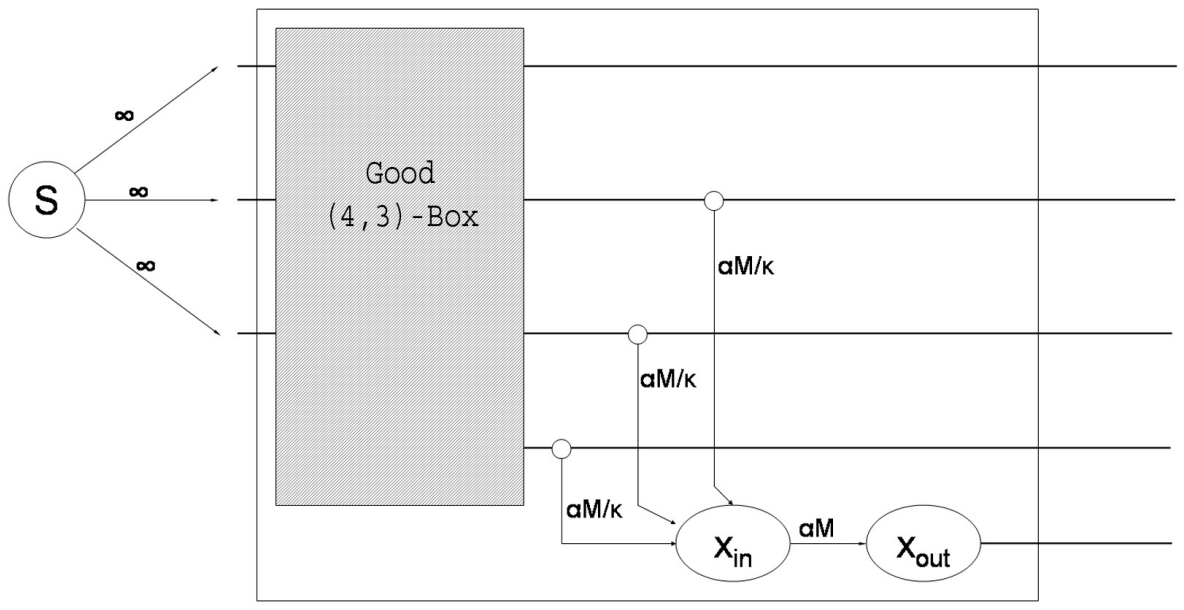

**Fig. 3.** Illustration of the inductive step. The internal box is good and we want to show that the external box is also good if the newcomer downloads $1/k\alpha\mathcal{M}$ from the existing nodes the big box is also good.

■

## V. Evaluation

In this section, we compare Regenerating Codes with other redundancy management schemes in the context of distributed storage systems. We follow the evaluation methodology of [20], which consists of a simple analytical model whose parameters are obtained from traces of node availability measured in several real distributed systems.

We begin in Section V-A with a discussion of node dynamics and the objectives relevant to distributed storage systems, namely reliability, bandwidth, and disk space. We introduce the model in Section V-B and estimate realistic values for its parameters in Section V-C. Section V-D contains the quantitative results of our evaluation. In Section V-E, we discuss qualitative tradeoffs between Regenerating Codes and other strategies, and how our results change the conclusion of [20] that erasure codes provide limited practical benefit.

### A. Background: node dynamics and objectives

In this section we introduce some background and terminology which is common to most of the work discussed in Section II-C.

We draw a distinction between *permanent* and *transient* node failures. A permanent failure, such as the permanent departure of a node from the system or a disk failure, results in loss of the data stored on the node. In contrast, data is preserved across a transient failure, such as a reboot or temporary network disconnection. We say that a node is *available* when its data can be retrieved across the network.

Distributed storage systems attempt to provide two types of reliability: availability and durability. A file is *available* when it can be reconstructed from the data stored on currently available nodes. A file's *durability* is maintained if it has not been lost due to permanent node failures: that is, it may be available at some point in the future. Both properties are desirable, but in this paper (as in [20]) we report results for availability only. Specifically, we will show *file unavailability*, the fraction of time that the file is not available.

As discussed in the introduction, achieving higher availability (or durability) implies a greater amount of redundancy, and hence uses more disk space, and more bandwidth to replace redundancy as nodes fail. Since bandwidth is generally considered a much more constrained resource than disk space in wide-area environments, we do not show the disk space used by the schemes we compare. However, disk usage would be proportional to bandwidth for all schemes we evaluate below, with the exception of OMMDS.

### B. Model

We use a model which is intended to capture the average-case bandwidth used to maintain a file in the system, and the resulting average availability of the file. With minor exceptions,[1]this model and the subsequent estimation of its parameters are equivalent to that of [20]. Although this evaluation methodology is a significant simplification of real storage systems, it allows us to compare directly with the conclusions of [20] as well as to calculate precise values for rare events.

The model has two key parameters, $f$ and $a$. First, we assume that in expectation a fraction $f$ of the nodes storing file data fail per unit time, causing data transfers to repair the lost redundancy. Second, we assume that at any given time while a node is storing data, the node is available with some probability $a$. Moreover, the model assumes that the event that a node is available is independent of the availability of all other nodes.

Under these assumptions, we can compute the expected availability and maintenance bandwidth of various redundancy schemes to maintain a file of $\mathcal{M}$ bytes. We make use of the fact that for all schemes except OMMDS (even Hybrid [20]), the amount of bandwidth used is equal to the amount of redundancy that had to be replaced, which is in expectation $f$ times the amount of storage used.

**Replication:** If we store $\mathcal{R}$ replicas of the file, then we store a total of $\mathcal{R} \cdot \mathcal{M}$ bytes, and in expectation we must replace $f \cdot \mathcal{R} \cdot \mathcal{M}$ bytes per unit time. The file is unavailable if no replica is available, which happens with probability $(1-a)^{\mathcal{R}}$.

**Ideal Erasure Codes:** For comparison, we show the bandwidth and availability of a hypothetical $(n,k)$ erasure code strategy which can "magically" create a new packet while transferring just $\mathcal{M}/k$ bytes (*i.e.*, the size of the packet). Setting $n = k \cdot \mathcal{R}$, this strategy sends $f \cdot \mathcal{R} \cdot \mathcal{M}$ bytes per unit time and has unavailability probability $U_{\text{ideal}}(n,k) := \sum_{i=0}^{k-1} \binom{n}{i} a^i (1-a)^{n-i}$.

**Hybrid:** If we store one full replica plus an $(n,k)$ erasure code where $n = k \cdot (\mathcal{R} - 1)$, then we again store $\mathcal{R} \cdot \mathcal{M}$ bytes in total, so we transfer $f \cdot \mathcal{R} \cdot \mathcal{M}$ bytes per unit time in expectation. The file is unavailable if the replica is unavailable *and* fewer than $k$ erasure-coded packets are available, which happens with probability $(1-a) \cdot U_{\text{ideal}}(n,k)$.

**OMMDS Codes:** A $(k,n)$ OMMDS Code with redundancy $\mathcal{R} = n/k$ stores $\mathcal{R}\mathcal{M}$ bytes in total, so $f \cdot \mathcal{R} \cdot \mathcal{M}$ bytes must be replaced per unit time. But replacing a fragment requires transferring over the network $\beta_{\text{OMMDS}} = (n-1)/(n-k)$ times the size of the fragment (see Section III-B), even in the most favorable case when newcomers connect to $n-1$ nodes to construct a new fragment. This results in $f \cdot \mathcal{R} \cdot \mathcal{M} \cdot \beta_{\text{OMMDS}}$ bytes sent per unit time, and unavailability $U_{\text{ideal}}(n,k)$.

**Regenerating Codes:** A $(k,n)$ Regenerating Code stores $\mathcal{M} \cdot n \cdot \beta_{\text{RC}}$ bytes in total (see Section IV). So in expectation $f \cdot \mathcal{M} \cdot n \cdot \beta_{\text{RC}}$ bytes are transfered per unit time, and the unavailability is again $U_{\text{ideal}}(n,k)$.

[1]In addition to evaluating a larger set of strategies and using a somewhat different set of traces, we count bandwidth cost due to permanent node failure only, rather than both failures and joins. Most designs [2], [27], [5] can avoid reacting to node joins. Additionally, we compute probabilities directly rather than using approximations to the binomial.

| Trace | Length (days) | Start date | Mean # nodes up | $f$ (fraction failed per day) | $a$ |
|---|---|---|---|---|---|
| PlanetLab | 527 | Jan. 2004 | 303 | 0.017 | 0.97 |
| Microsoft PCs | 35 | Jul. 6, 1999 | 41970 | 0.038 | 0.91 |
| Skype | 25 | Sept. 12, 2005 | 710 | 0.12 | 0.65 |
| Gnutella | 2.5 | May, 2001 | 1846 | 0.30 | 0.38 |

**TABLE I:** The availability traces used in this paper.

### C. Estimating $f$ and $a$

In this section we describe how we estimate $f$, the fraction of nodes that permanently fail per unit time, and $a$, the mean node availability, based on traces of node availability in several distributed systems.

We use four traces of node availability with widely varying characteristics, summarized in Table I. The **PlanetLab All Pairs Ping [24]** trace is based on pings sent every $15$ minutes between all pairs of $200$-$400$ nodes in PlanetLab, a stable, managed network research testbed. We consider a node to be up in one $15$-minute interval when at least half of the pings sent to it in that interval succeeded. In a number of periods, all or nearly all PlanetLab nodes were down, most likely due to planned system upgrades or measurement errors. To exclude these cases, we "cleaned" the trace as follows: for each period of downtime at a particular node, we remove that period (i.e. we consider the node up during that interval) when the average number of nodes up during that period is less than half the average number of nodes up over all time. The **Microsoft PCs [4]** trace is derived from hourly pings to desktop PCs within Microsoft Corporation. The **Skype superpeers [12]** trace is based on application-level pings at 30-minute intervals to nodes in the Skype superpeer network, which may approximate the behavior of a set of well-provisioned endhosts, since superpeers may be selected in part based on bandwidth availability [12]. Finally, the trace of **Gnutella peers [22]** is based on application-level pings to ordinary Gnutella peers at 7-minute intervals.

We next describe how we derive $f$ and $a$ from these traces. It is of key importance for the storage system to distinguish between permanent and transient failures (defined in Section V-A), since only the former requires bandwidth-intensive replacement of lost redundancy. Most systems use a *timeout* heuristic: when a node has not responded to network-level probes after some period of time $t$, it is considered to have failed permanently. To approximate a storage system's behavior, we use the same heuristic. Node availability $a$ is then calculated as the mean (over time) fraction of nodes which were available among those which were not considered permanently failed at that time.

The resulting values of $f$ and $a$ appear in Table I, where we have fixed the timeout $t$ at $1$ day. Longer timeouts reduce overall bandwidth costs [20], [5], but begin to impact durability [5] and are more likely to produce artificial effects in the short ($2.5$-day) Gnutella trace.

We emphasize that the procedure described above only provides an estimate of $f$ and $a$ which may be biased in several ways. Some designs [5] reincorporate data on nodes which return after transient failures which were longer than the timeout $t$, which would reduce $f$. Additionally, even placing files on uniform-random nodes results in selecting nodes that are more available [25] and less prone to failure [11] than the average node. Finally, we have not accounted for the time needed to transfer data onto a node, during which it is effectively unavailable. However, we consider it unlikely that these biases would impact our main results since we are primarily concerned with the *relative* performance of the strategies we compare.

### D. Quantitative results

Figure 4 shows the tradeoff between mean unavailability and mean maintenance bandwidth in each of the strategies of Section V-B using the values of $f$ and $a$ from Section V-C and $k = 7$. Figure 5 shows the tradeoff for $k = 14$. Points in the tradeoff space are produced by varying the redundancy factor $\mathcal{R}$.

We show OMMDS in the two Skype plots where the trends are clearly visible. To reduce clutter we omit similar results for the other traces. In all cases, OMMDS obtains worse points in the tradeoff space than Hybrid.

In the more stable environments, Regenerating Codes obtain a substatantial benefit over Hybrid strategy. For example, in the PlanetLab trace with $k = 7$, RC has about $25\%$ lower bandwidth for the same availability, or more than $3$ orders of magnitude lower unavailability with the same bandwidth. The difference is even greater for $k = 14$.

RC's reduction in bandwidth compared with Hybrid diminishes as the environment becomes less stable; in the most extreme case of the Gnutella trace, RC can actually be very slightly *worse*. The reason can be seen by comparing the two schemes with Ideal Erasure Codes. For fixed $k$ and $n$, both RC and Hybrid have roughly the same availability (Hybrid is slightly better due to the extra replica). However, in terms of bandwidth as we scale $n$, RC has a small *constant factor* overhead compared with Ideal Erasure codes, while Hybrid has a rather large but only *additive* overhead due to the single extra replica. For large enough $n$, such as is necessary in Gnutella, the additive overhead wins out.

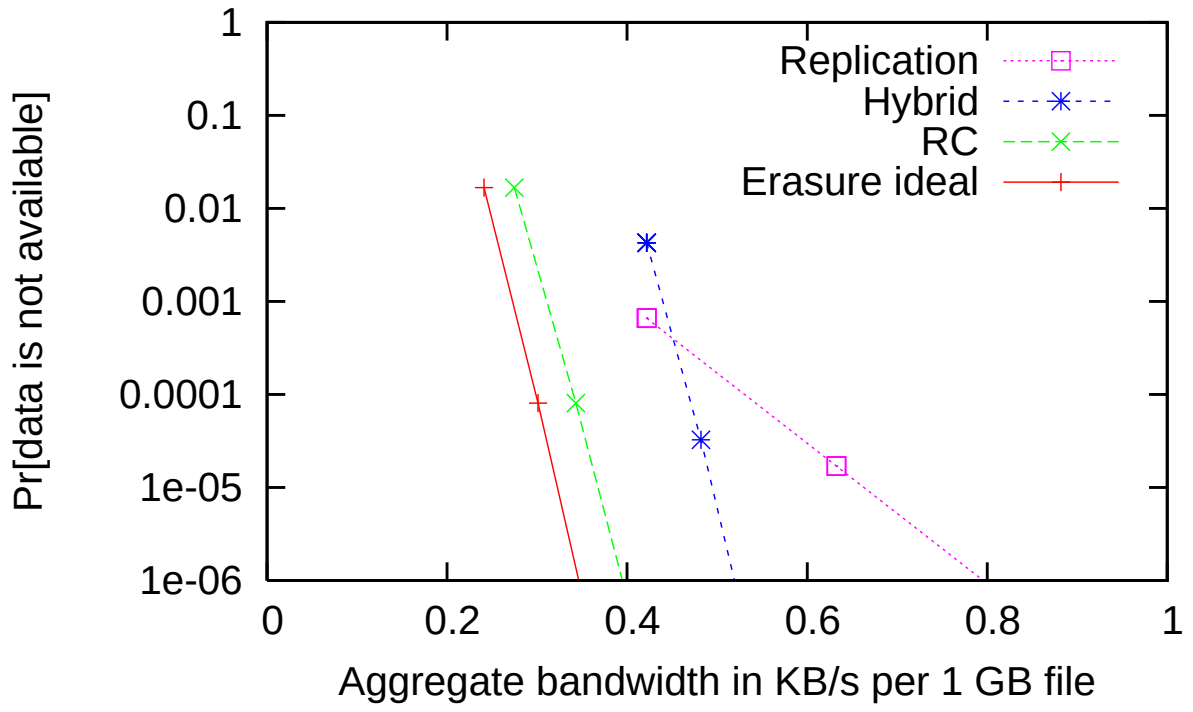

(a) PlanetLab trace

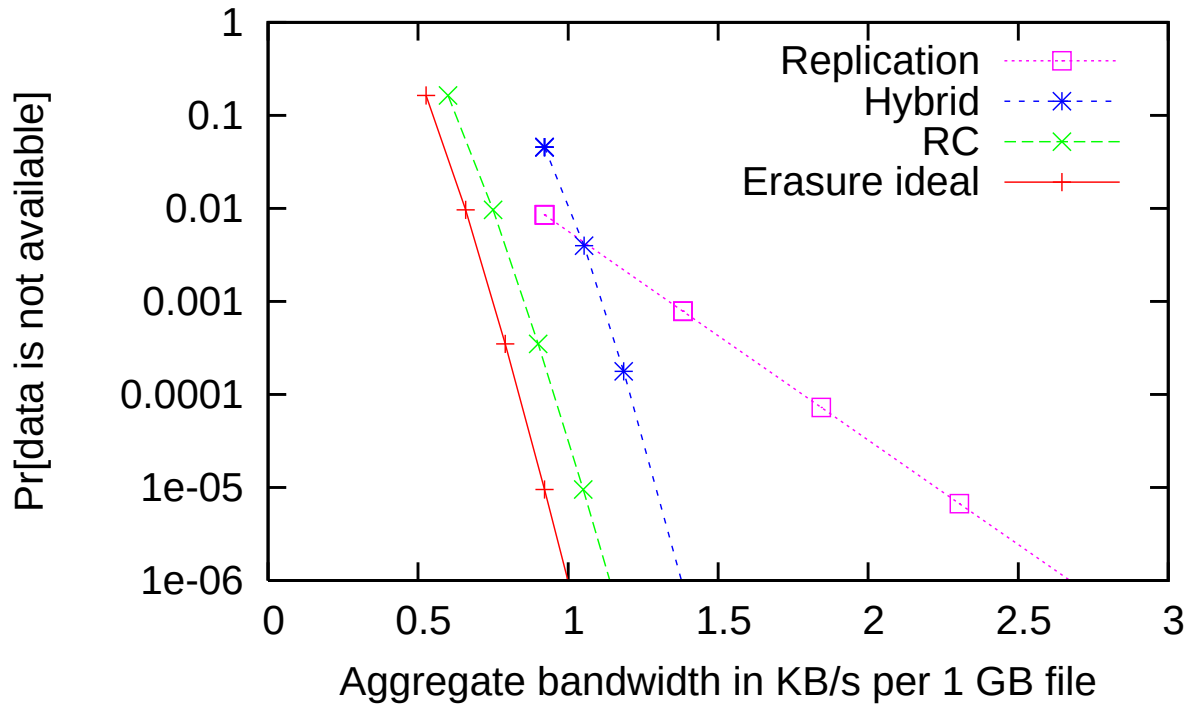

(b) Microsoft PCs trace

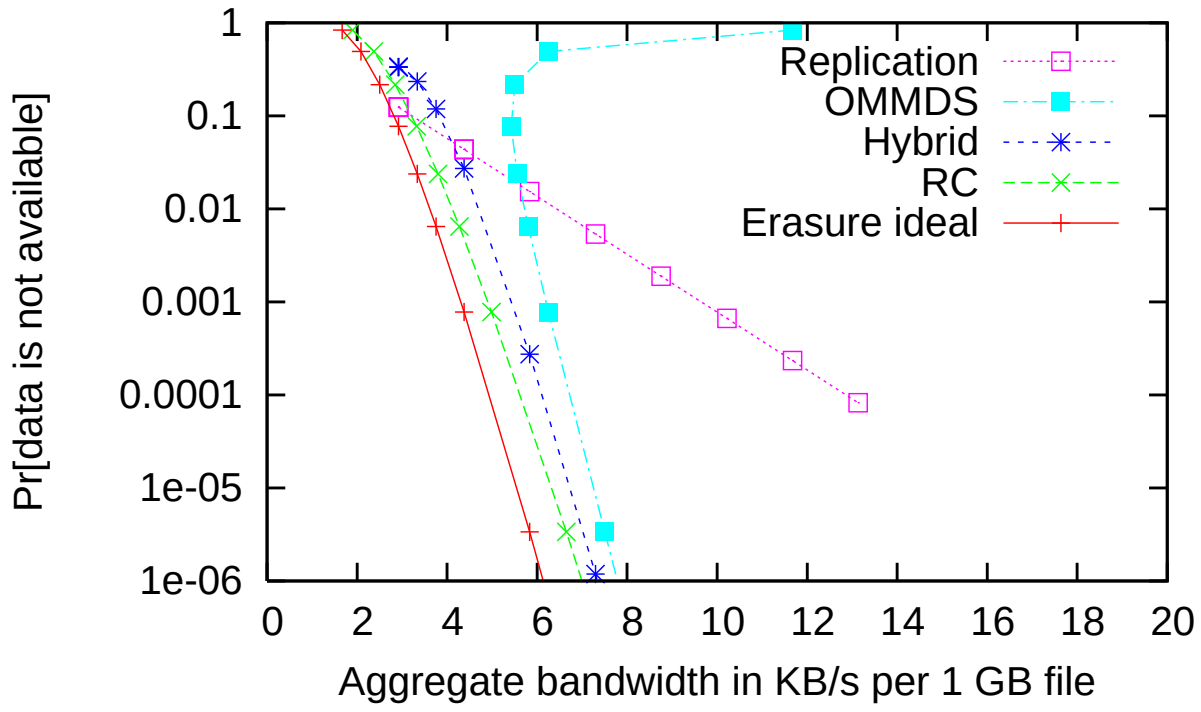

(c) Skype superpeers trace

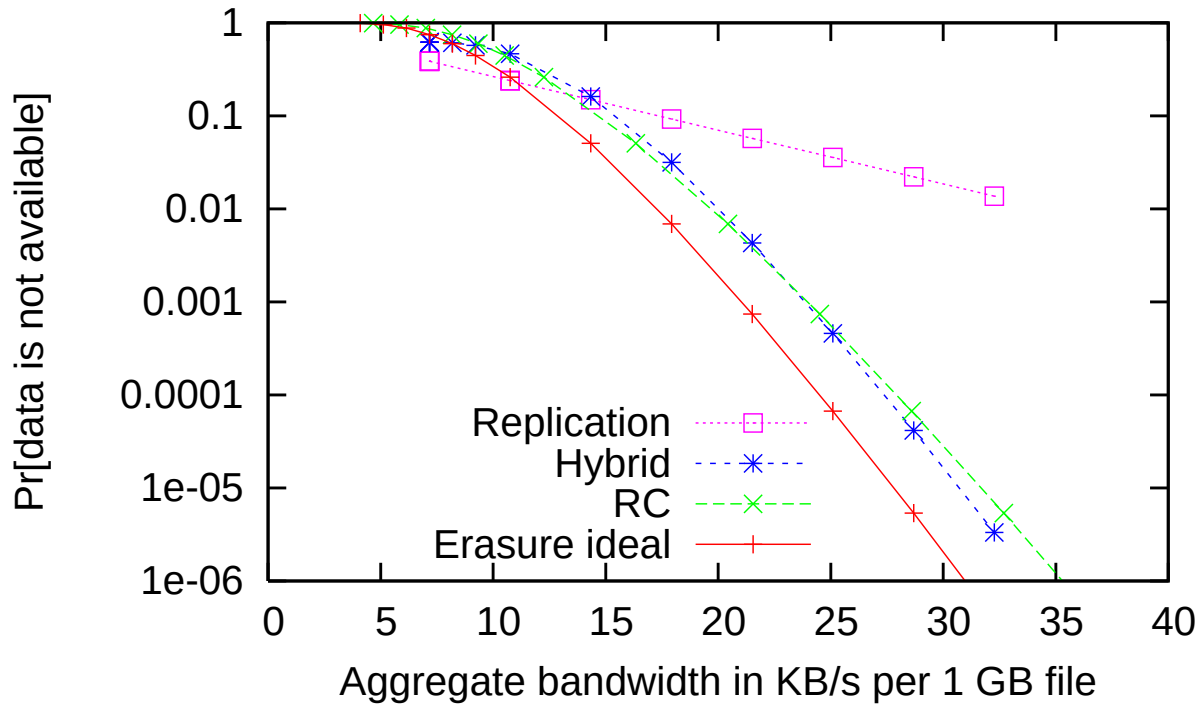

(d) Gnutella peers trace

**Fig. 4:** Availability-bandwidth tradeoff for $k = 7$ with parameters derived from each of the traces.

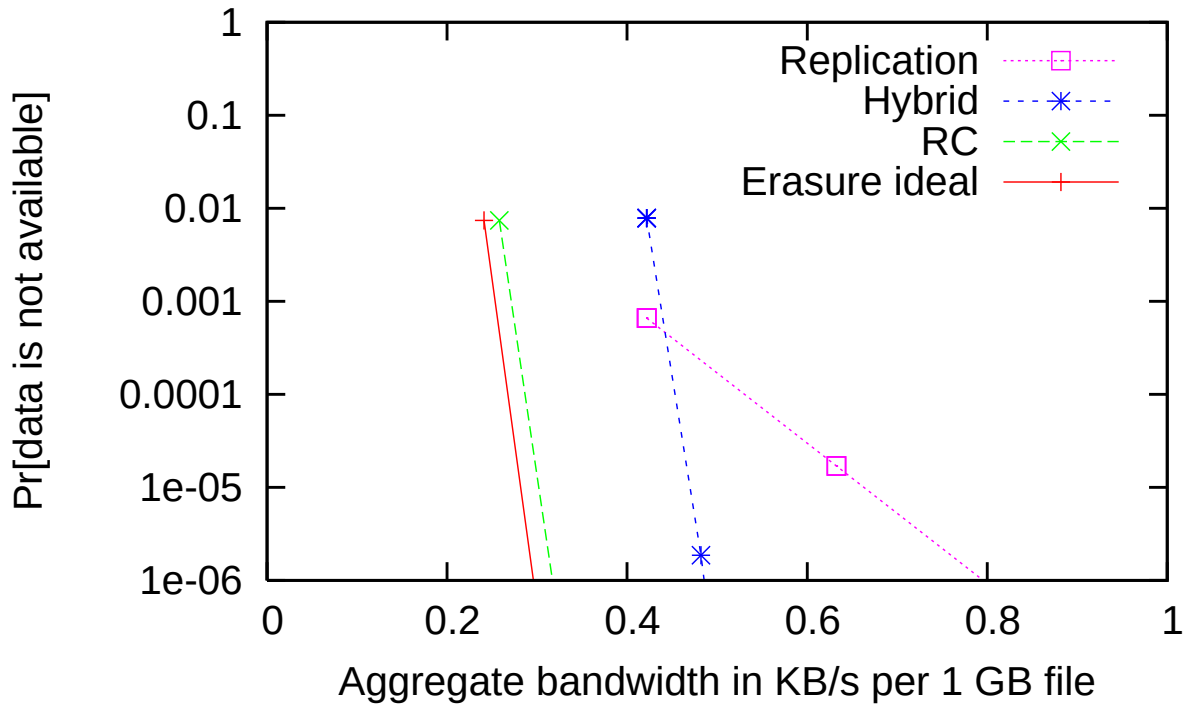

(a) PlanetLab trace

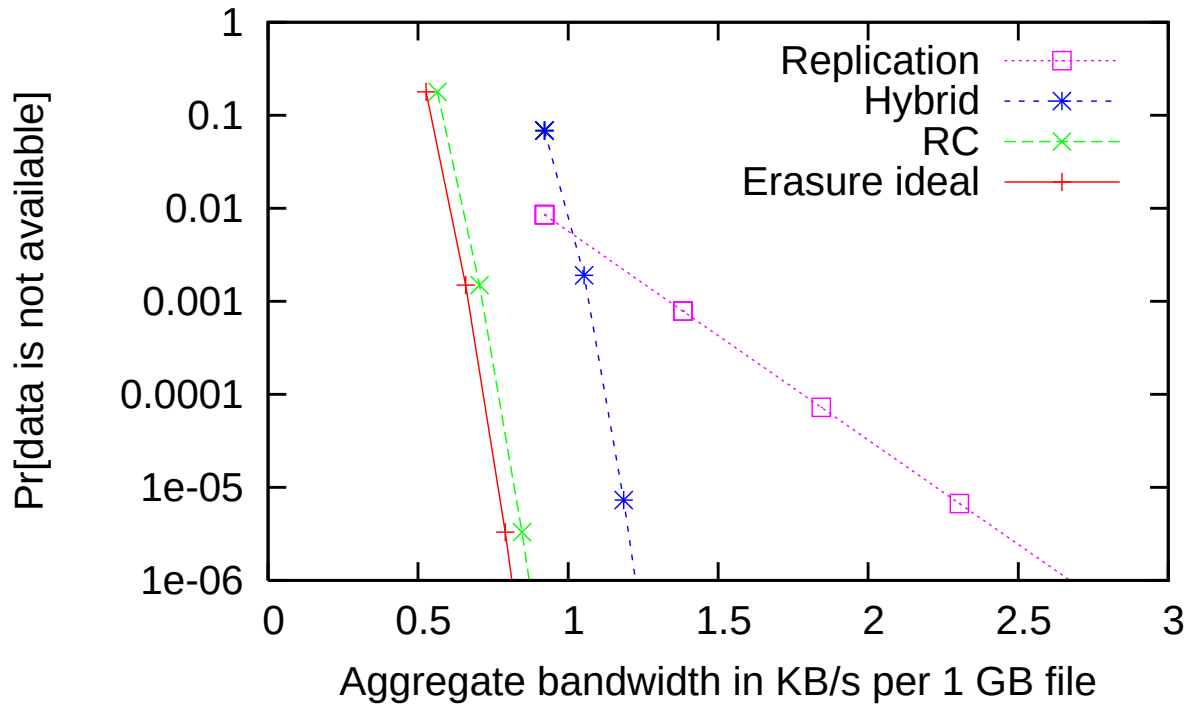

(b) Microsoft PCs trace

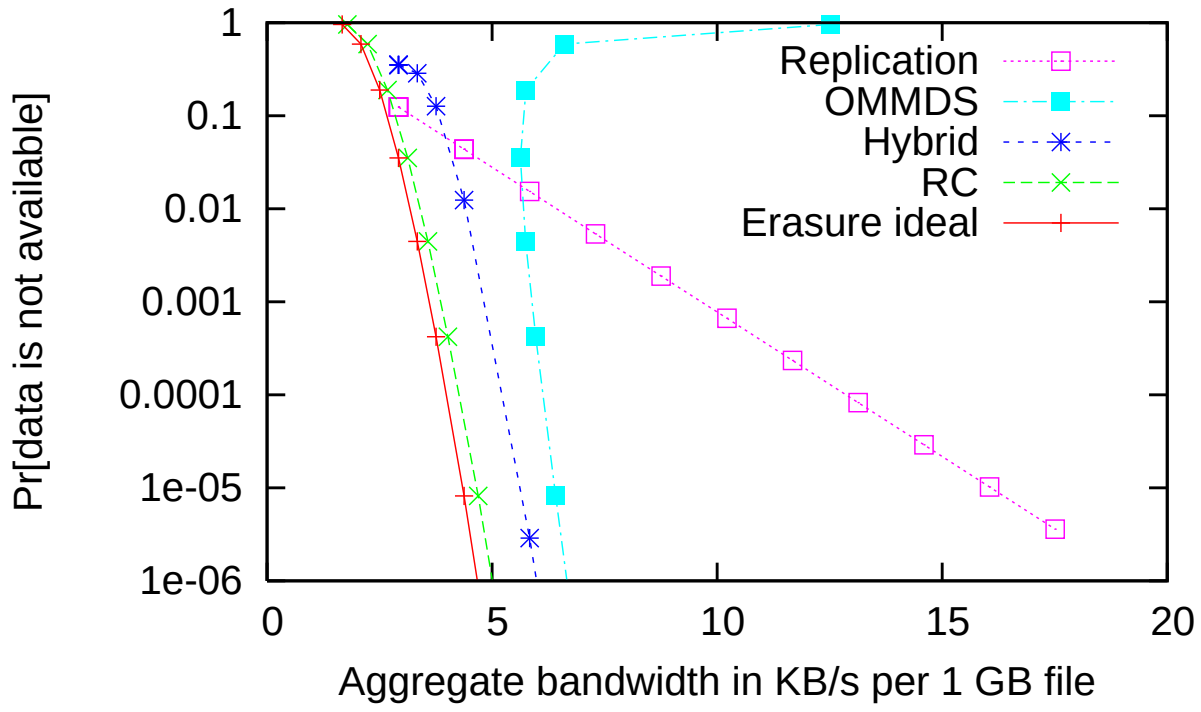

(c) Skype superpeers trace

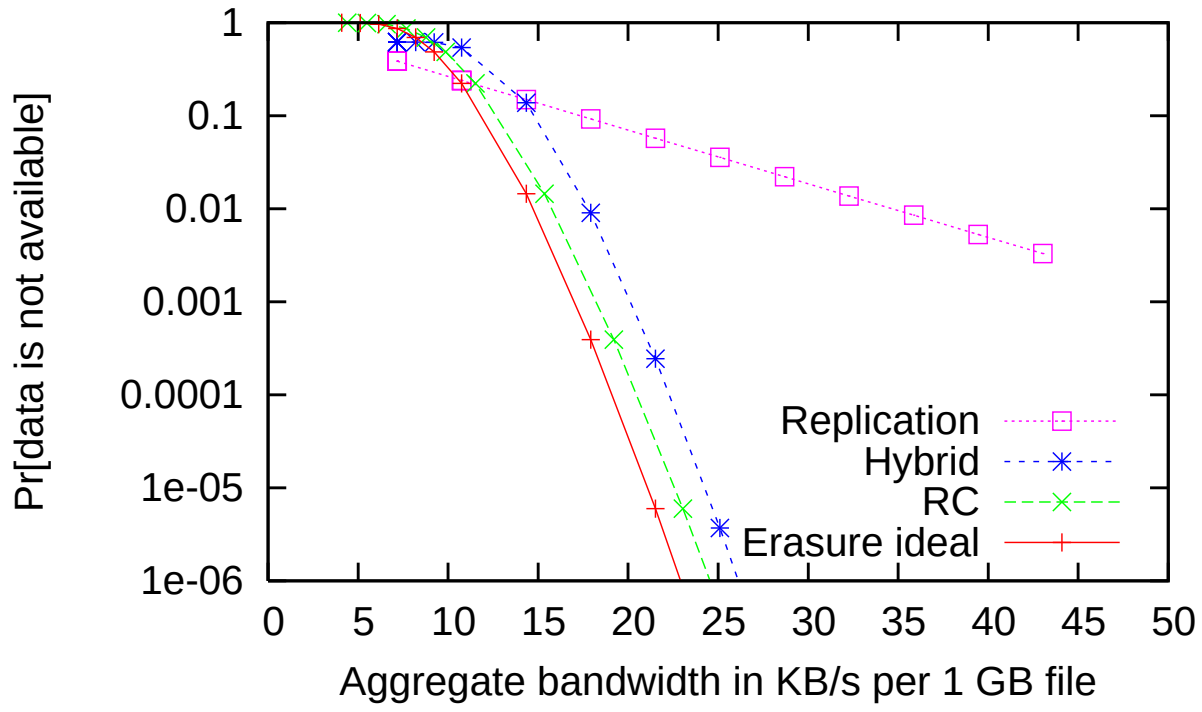

(d) Gnutella peers trace

**Fig. 5:** Availability-bandwidth tradeoff for $k = 14$ with parameters derived from each of the traces.

This is unlikely to make Hybrid a lower-bandwidth choice, for two reasons. First, as demonstrated by Figure 5, a larger value of $k$ diminishes RC's overhead sufficiently that it is better than Hybrid in all cases that we tested. Second, as discussed in [20], more stable environments are the more likely deployment scenario in any case. Taking numbers from the Gnutella trace with a target unavailability of $0.01$, nodes are unlikely to want to use $20$ KB/sec of bandwidth (a significant fraction of typical endhost uplink bandwidth) just to reliably store a $1$ GB file (a very small fraction of today's hard disks). In PlanetLab, the same $1$ GB can be maintained with $100\times$ lower unavailability using about $58\times$ less bandwidth.

### E. Qualitative comparison

In this section we discuss two questions: First, is RC an overall win over Hybrid? Second, do our results affect the conclusion of Rodrigues and Liskov [20] that erasure codes offer too little improvement in bandwidth use to clearly offset the added complexity that they add to the system?

The results of Section V-D suggest that in practical scenarios RC provides a significant reduction in maintenance bandwidth over Hybrid, as well as simplifying system architecture since only one type of redundancy needs to be maintained. This addresses the two principal disadvantages of using erasure codes discussed in [20].

However, RC still has some drawbacks. First, constructing a new packet, or reconstructing the entire file, requires communcation with $k$ nodes rather than one (in Hybrid, the node holding the single replica). This adds overhead that could be significant for sufficiently small files or sufficiently large $k$. Perhaps more importantly, as discussed in Section IV, there is a factor $\beta_{\mathrm{RC}}$ increase in total data transferred to *read* the file, roughly 14% for $k = 7$ but diminishing to 7.1% for $k = 14$ and 3.1% for $k = 32$ (see Figure 2). Thus, if the frequency that a file is read is sufficiently high and $k$ is sufficiently small, this inefficiency could overwhelm the reduction in maintenance bandwidth.

If the target application is archival storage or backup, files are likely to be large and infrequently read. We believe this is one case in which RC is likely to be a significant win over both Hybrid and replication.